\documentclass[11pt, epsf]{article}
\usepackage{epsfig}
\def\ga{\mathrel{\raise.3ex\hbox{$>$\kern-.75em\lower1ex\hbox{$\sim$}}}}
\def\la{\mathrel{\raise.3ex\hbox{$<$\kern-.75em\lower1ex\hbox{$\sim$}}}}

\def\I_M{{I_{\scriptscriptstyle M\times M}}}

\begin{document}

\thispagestyle{empty}
\rightline{IP/BBSR/2004-06-17}
\rightline{{\tt hep-th/0406169}}

\vskip 2cm \centerline{ \Large \bf Born-Infeld black holes in the presence 
of a }
\vskip .2cm \centerline{ \Large \bf cosmological constant}

\vskip .2cm

\vskip 1.2cm

\centerline{ \bf Tanay  Kr. Dey}
\vskip 10mm \centerline{ \it Institute of Physics, 
Bhubaneswar-751005, India} 
\vskip 1.2cm

\centerline{\tt tanay@iopb.res.in}

\vskip 1.2cm

\begin{quote}
\centerline{\bf Abstract}
\vskip 1.2cm

We construct asymptotically anti-deSitter (and deSitter) black hole 
solutions of 
Einstein-Born-Infeld theory in arbitrary dimension. We critically analyse 
their geometries and discuss their  thermodynamic properties.  
\end{quote}

\newpage
\setcounter{footnote}{0}
\section{Introduction}
\noindent By now, there are substantial evidences which suggests 
that 
the type IIB string theory on $AdS_5\times S_5$ is dual to four 
dimensional $N = 4$ super Yang-Mills theory \cite{review}. At high 
temperature, the 
thermal state of the string theory is described by an asymptotically AdS 
black 
hole. Therefore, qualitatively the properties of thermal super Yang-Mills 
can be understood by studying the black hole geometry. However, due to 
the fact that the 
bulk supergravity describes gauge theory at strong coupling, the 
quantitative understanding becomes difficult. Nevertheless, many attempts 
were made along this direction (see for example \cite{gkt}, \cite{li}, 
\cite{lindstrom}, 
\cite{mozo}, \cite{myers}). 
By going higher in the bulk coupling, we make the dual boundary theory 
weaker. For example, one may wonder if we can atleast qualitatively 
understand the behaviour of the boundary Yang-Mills theory as we perturb 
the black hole geometry by turning on the higher derivative curvature 
terms in the bulk action. In fact, such an analysis was carried out in 
\cite{nod} by constructing bulk anti-deSitter black holes 
in the presence of certain higher curvature terms in the supergravity 
action. Furthermore, black holes of higher curvature gravity were 
constructed where the electromagnetic coupling was turned on \cite{banados} 
\footnote{Under 
IIB 
string compactification, the electromagnetic field appear when we take 
the compact space to be a spinning sphere.}. In general, beside the 
curvature terms, one would also expect higher derivative gauge field 
contributions to the supergravity actions. How does the boundary theory 
respond when we incorporate such corrections? As a first step to analyse 
such an issue, in this paper, we study
the effect of adding higher derivative gauge field  terms on the bulk 
adS (dS) black hole geometry. This is done by explicitly constructing 
black hole solutions of the supergravity action coupled to a Born-Infeld 
gauge field in arbitrary dimension in the presence of a cosmological 
constant.
This action not only incorporates the higher order gauge 
field corrections to the Einstein-Maxwell gravity in the presence of a 
cosmological constant, but also allows us to find exact black hole 
solutions. Note that in general the gravity action may have both 
higher order curvature terms as well as the higher derivative terms due to 
gauge fields. We have not analysed here the black holes in these.

\noindent In the remaining part of the paper, we first construct  black 
holes in $(n+1)$ dimensional Einstein-Born-Infeld theory in the presence of 
a negative or positive cosmological constant\footnote{ In three 
and four 
dimensions the solutions were constructed in \cite{cg} and \cite{fk} 
respectively. However, their 
thermodynamical behaviour was not analysed.}. Next, we study the 
thermodynamics of these holes. Here, in particular, we check that these
black holes follow first law of thermodynamics. Then, by calculating 
specific heat at fixed charge, we show that for a certain range of 
parameters these black holes are stable. Expression of free energy at 
fixed charge is also calculated for these holes. Furthermore, by expanding 
our solutions around Reissner-Nordstrom anti deSitter (RNadS) black holes
\cite{emparan}, we find out the effect of higher order gauge field 
corrections to the geometry.
\section{{\hspace{-.5cm} AdS and dS Black hole solutions of Einstein-Born-Infeld theory }}

\noindent In this section, we will explicitly construct the black hole 
solutions of 
Einstein-Born-Infeld action in $(n+1)$ dimension in the presence of a 
negative (or positive) cosmological constant  $\Lambda$. The action has 
the form
\begin{equation}
{ S} = {\int {d^{n+1}x}\sqrt{-g}\Big[\frac{(R-2\Lambda)}{16\pi G} + 
L(F)\Big]}, 
\label{}
\end{equation}
where $ L(F)$  is given by
\begin{equation}
{ L(F)} = {4{\beta}^2\Big (1-\sqrt{1+\frac{F^{\mu \nu} 
F_{\mu \nu}}{2{\beta}^2}}\Big)}.
\label{}
\end{equation}
The constant $\beta$  is the Born-Infeld parameter and has
the dimension of mass. In the limit $\beta \rightarrow \infty $  $ L(F)$ 
reduces to the standard Maxwell form
\begin{equation}
{ L(F)} = {- F^{\mu \nu} F_{\mu \nu}} +{\cal{O}}(F^4).
\label{}
\end{equation}
The black holes in this limit were constructed in \cite{emparan} and their 
thermodynamics and phase structures were studied in \cite{emp}, 
\cite{emparan}. 
For simplicity, in this paper, we will work 
with the convention that $16 \pi G =1$, where $G$ is the Newton's 
constant.

\noindent By varying the action with respect to the gauge field $ A_\mu
$ and the metric  $ g_{\mu \nu} $, we get the corresponding equations
of motion. These are respectively 
\begin{equation}
{{ \bigtriangledown}_\mu \Big(\frac{ F^{\mu \nu}}
{\sqrt{1+\frac{F^2}{2{\beta}^2}}}\Big)} ={0},
\label{}
\end{equation}
and
\begin{equation}
R_{\mu \nu}+\frac{2}{n-1} g_{\mu \nu}\Lambda = \frac{1}{n-1}
  g_{\mu \nu}L+ \frac{2}{n-1}{\frac{ g_{\mu \nu}
      F^2}{\sqrt{1+\frac{F^2}{2{\beta}^2}}} -
\frac{2 F_{\alpha \mu}{F^{\alpha}}_\nu}{\sqrt{1+\frac{F^2}{2{\beta}^2}}}}.
\label{}
\end{equation}
In order to solve the equations of motion, we use the metric 
ansatz
\begin{equation}
{ds^2} = {-V(r)dt^2 + \frac{dr^2}{V(r)} + r^2 d{{\Omega}^2}_{n-1}},
\label{}
\end{equation}
where, $ d{{\Omega}^2}_{n-1} $ denotes the metric of an unit $ (n-1)$ sphere.
$V(r)$ is an unknown function of $r$ which we will determine shortly.
First of all, a class of solution of equation (4) can immediately be 
written down where all the
components of $F^{\mu\nu}$ are zero except $F^{rt}$. It is given by
\begin{equation}
{F^{rt}} = {\frac{\sqrt{(n-1)(n-2)}\beta q}{\sqrt{2{\beta}^2
        {r}^{2n-2}+(n-1)(n-2)q^2}}}.
\label{}
\end{equation}
Here $ q $ is an integration constant and is related to the 
electromagnetic charge. This can be concluded from the behaviour 
of $F^{rt}$ in the large $\beta$ limit as   $F^{rt} \sim \frac{q}{r^{n-1}}$. 
We notice that the electric field is 
finite at $r =0$. This is expected in Born-Infeld theories.
Now, parametrising $ \Lambda = -\frac{n(n-1)}{2l^2} $,
equation (6) can easily be solved as
\begin{equation}
 V(r)= 1-\frac{m}{r^{n-2}}+\Big[\frac{4{\beta}^2}{n(n-1)}+\frac{1}{l^2}\Big]r^2
 -{{\frac{2{\sqrt2}\beta}{(n-1)r^{n-2}}{\int{\sqrt{2{\beta}^2 
        r^{2n-2}+(n-1)(n-2)q^2}}dr}}}.
\label{}
\end{equation}
The integral can be done in terms of hypergeometric function and can be
written in a compact form. The result is
\begin{eqnarray}
 V(r)&=& 
1-\frac{m}{r^{n-2}}+\Big[\frac{4{\beta}^2}{n(n-1)}+\frac{1}{l^2}\Big]r^2-{\frac {2 {\sqrt{2}}{\beta}}{n(n-1)r^{n-3}}}{\sqrt {2 \beta^2 r^{2n-2}+{(n-1)(n-2)q^2}}}\nonumber \\
&+&{\frac{2(n-1)q^2}{nr^{2n-4}}}{}_2F_1\Big[ {{n-2}\over{2n-2}}, {1\over 2}, {{3n-4}\over{2n-2}}, 
-{\frac{(n-1)(n-2)q^2}{2 \beta^2 r^{2n-2}}}\Big].
\end{eqnarray}
In the above expression, $m$ appears as an integration constant 
and is related to the ADM mass of the
configuration.
It can be checked that for $n=3$, it reduces to the solution of \cite{fk}. 
Also, in the $l \rightarrow \infty$ limit (that is for $\Lambda =0$), this 
solution reproduces correctly the asymptotically flat Born-Infeld black 
hole (see for example \cite{rasheed}).

\noindent Using the fact that $_2F_1(a,b,c,z)$ has a convergent series 
expansion for $|z| <1$, we can find the behaviour of the metric for large 
$r$. This is given by
\begin{equation}
V(r) = 1 - {m\over{r^{n-2}}} + {q^2\over{r^{2n-4}}} + {r^2\over {l^2}}
- {(n-1)(n-2)^2 q^4\over{8 \beta^2 (3n -4)r^{4n-6}}}, 
\end{equation}
Note that in the $\beta \rightarrow \infty$, it has the form of 
Reissner-Nordstrom adS black hole \cite{emparan}.
The last term in the right hand side of the above expression is the 
leading Born-Infeld correction to the RNadS black hole in the large 
$\beta$ limit. From the asymptotic behaviour, we see that $m$ is related 
to the mass of the configuration. In particular, in our convention, the 
ADM mass $M$ is 
\begin{equation}
M = (n-1) \omega_{n-1} m,
\label{mas}
\end{equation}
where $\omega_{n-1}$ is the volume of the unit $(n-1)$ sphere.
More interesting is the behaviour of $V(r)$ close to the 
origin where 
\begin{equation}
V(r) =1 - {{m-A}\over{r^{n-2}}}-\Big[\frac{2 c\beta }{n}- B
(2n-1)q\Big]{\frac{q}{r^{n-3}}}+\Big[\frac{4{\beta}^2}{n(n-1)} +
{1\over {l^2}}\Big]r^2 -\Big[\frac{2c{\beta}}{n}
+B\Big]{\frac{{\beta}^2 r^{n+1}}{(n-1)(n-2)}},
\end{equation}
where 
\begin{equation}
 A = {\frac{2(n-1) q^2 }{n \sqrt{\pi}}}{\Big\lbrace{\frac{2
{\beta}^2} 
{(n-1)(n-2)q^2}}\Big\rbrace}^{\frac{n-2}{2n-2}} 
{\Gamma{\Big[}\frac{3n-4}{2n-2}{\Big]}\Gamma{\Big[}\frac{1}{2n-2}{\Big]}},
\label{cnst}
\end{equation}
and
$$ c = \sqrt{\frac{2(n-2)}{(n-1)}} {~~~}{\rm and}{~~~} B = {\frac{4 
\beta}{cn(2n-1)q}}{\frac{\Gamma{\Big[}\frac{3n-4}{2n-2}{\Big]}
\Gamma{\Big[}\frac{-1}{2n-2}{\Big]}}{\Gamma
{\Big[}\frac{n-2}{2n-2}{\Big]}\Gamma{\Big[}\frac{2n-3}{2n-2}{\Big]}}}.$$
From the above expression, we see that for generic values of $n \ge 3$, 
the metric has a curvature singularity at $r=0$. However, for $n=3$ and 
for $m=A$, the metric is regular at $r=0$. We now proceed to analyse if 
this singularity is hidden behind a horizon.
The horizons correspond to the locations where 
$V(r) = 0$. Though we are unable to solve this equation analytically, we 
first plot, in figure 1, the function $V(r)$ for some  different values of $m$ 
and for $n=4$. In this figure, the other parameters such as $l, \beta$ are 
kept fixed. First of all, let us note that there can be one or two 
horizons depending on the value of $m$. Furthermore, for certain choices 
of $m$ there can be no horizon, leading to a naked singularity at the 
origin. To have further understanding on the nature of the horizons, we 
plot in figure 2, the mass as a function of the horizon radius.
\begin{figure}[ht]
\epsfxsize=8cm
\centerline{\epsfbox{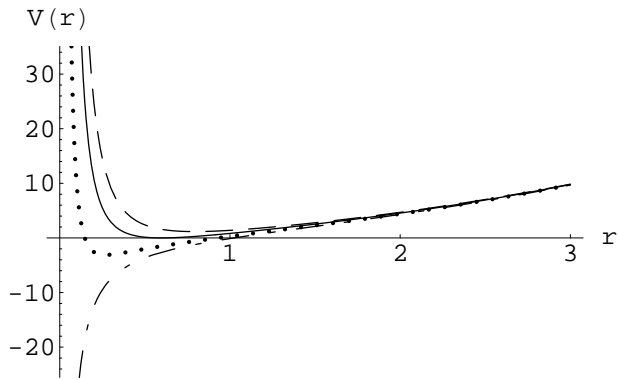}}
\caption{{\small{The metric function $V(r)$ as a 
function of $r$ for $n=4, \beta =1, q=1, l=1$. The dashed line, solid 
line, 
dotted line and dash-dotted line are for $m=1.5, 2.062, 2.6, 3$ respectively.}} }
\end{figure}
The mass parameter of the hole can be expressed in
terms of horizon radius $(r_+)$ as
\begin{eqnarray}
m&=&r_+^{n-2}  +\Big[ {4 {\beta}^2\over{n(n-1)}}+ \frac{1}{l^2}\Big]r_+^n 
-{\frac {2 {\sqrt{2}}{\beta}r_+}{n(n-1)}}{\sqrt {2 \beta^2 r_+^{2n-2}+{(n-1)(n-2)q^2}}}\nonumber \\
&+&{\frac{2(n-1)q^2}{nr_+^{n-2}}}{}_2F_1\Big[ {{n-2}\over{2n-2}}, {1\over 2}, {{3n-4}\over{2n-2}}, 
-{\frac{(n-1)(n-2)q^2}{2 \beta^2 r_+^{2n-2}}}\Big].
\end{eqnarray}
In figure 2, $m(r_+)$ is shown for different values of $\beta$. For a 
given $\beta$, the number of horizons depend clearly upon the choice of 
$m$. If we focus our attention to the solid line ($\beta = 5$), we see 
that upto certain $m$, there are two horizons. As we decrease the mass 
further, the two horizons meet.
\begin{figure}[ht]
\epsfxsize=8cm
\centerline{\epsfbox{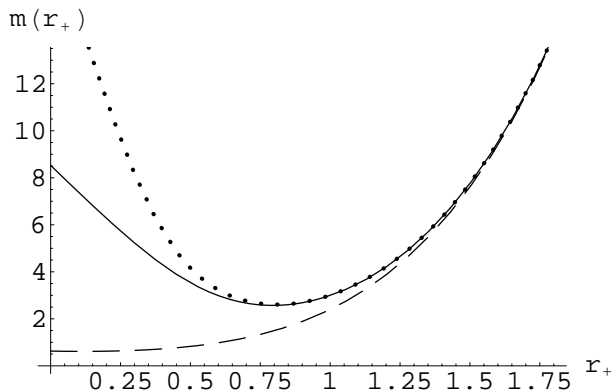}}
\caption{{\small{The mass $m$ as a function of $r_+$ for adS Born-Infeld holes
for $n=4, l=1, q=1, 
\beta =16$ (dotted line) ,$5$ (solid line) ,$0.1$ (dashed line).}}}
\end{figure}
\noindent We then call the black hole  
extremal. It is easy to find out from $V(r)$ that this happens when $r_+$ 
satisfies the following condition:
\begin{equation}
(n-2)r_+^{n-3}  +\Big[ \frac{4 {\beta}^2}{n-1}+ \frac{n}{l^2}\Big]r_+^{n-1}-{\frac{2\sqrt{2}\beta}{n-1}}{\sqrt{2 {\beta}^2
  r_+^{2n-2} + (n-1)(n-2)q^2}}= 0.
\label{extr}
\end{equation}
For $ \beta =0.1 $ there is only a single horizon, as equation (15) does not
have a real solution for $ r_+ $ and hence extremality condition can
not be satisfied.
As we will see later that when this condition is satisfied, the 
temperature of the black hole vanishes. We also notice that the behaviour 
of $V(r)$ crucially depends on the ratio $m/A$, where $A$ is defined in
(\ref{cnst}). If $m/A \ge 1$, $V(r)$ behaves like that of a Schwarzschild 
black hole close to the origin. On the other hand, for $m/A < 1$, 
behaviour of $V(r)$ is more like the Reisssner-Nordstrom one.\\ \\
\noindent From equation (7), we can also 
calculate the gauge field associated with this configuration. It is given 
by
\begin{eqnarray}
A_{t}={\frac{1}{c}}{ \frac{q}{r^{n-2}}~{}_2F_1\Big[ {{n-2}\over{2n-2}}, 
{1\over 2}, {3n-4\over{2n-2}}, 
-{\frac{(n-1)(n-2)q^2}{2{\beta}^2 r^{2n-2}}}\Big]}-\Phi,
\label{gaug}
\end{eqnarray}  
where $q$ is related to the black hole charge $Q$ via
 $$ {Q}={2\sqrt{2(n-1)(n-2)}{\omega}_{n-1}q}{~}. $$
In equation (\ref{gaug}), $\Phi$ is the gauge potential. We will
choose $\Phi$ in such a way that $A_t$ is zero at the horizon. This gives
\begin{equation}
\Phi = {\frac{1}{c}}{\frac{q}{{r_+}^{n-2}}~{}_2F_1\Big[ 
{{n-2}\over{2n-2}}, {1\over 2}, {3n-4\over{2n-2}}, 
-{\frac{(n-1)(n-2)q^2}{2{\beta}^2 {r_+}^{2n-2}}}\Big]}.
\end{equation}
Behaviour of $\Phi$ as a function of the horizon size for $n=4,q=4, 
l=\beta 
=1$ is shown in figure 3. Notice that $\Phi$ is finite even when $r_+ =0$.
\begin{figure}[ht]
\epsfxsize=9cm
\centerline{\epsfbox{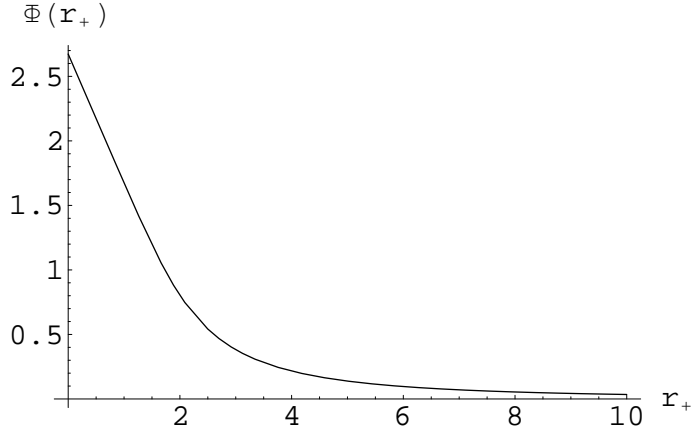}}
\caption{{\small{The potential $\Phi$ as a function of 
$r_+$ for $n=4, l=1, {\beta}=1 ,q=4$. Note that $\Phi$ is finite for $r_+ 
=0$.}}}                             
\end{figure}

\noindent We would now like to make some brief comments on the Born-Infeld 
deSitter 
black holes. These are the solutions in the presence of a positive cosmological 
constant. This can be found from the earlier expression of $V(r)$ by 
replacing $l^2$ by $-l^2$. More explicitly,
\begin{eqnarray}
 V(r)&=& 
1-\frac{m}{r^{n-2}}+\Big[\frac{4{\beta}^2}{n(n-1)}-\frac{1}{l^2}\Big]r^2-{\frac {2 {\sqrt{2}}{\beta}}{n(n-1)r^{n-3}}}{\sqrt {2 \beta^2 r^{2n-2}+{(n-1)(n-2)q^2}}}\nonumber \\
&+&{\frac{2(n-1)q^2}{nr^{2n-4}}}{}_2F_1\Big[ {{n-2}\over{2n-2}}, {1\over 2}, {{3n-4}\over{2n-2}}, 
-{\frac{(n-1)(n-2)q^2}{2 \beta^2 r^{2n-2}}}\Big].
\end{eqnarray}
This metric, in the asymptotic region, goes to the Reisnner-Nordstrom 
deSitter black holes with a $\beta$ dependent Born-Infeld correction. 
While near $r=0$,  $V(r)$ behaves similar to equation (12) with $l^2 
\rightarrow -l^2$. Therefore, the singularity structure of this solution 
is the same as the previous one. Now, turning our attention to the nature 
of the horizon, we find that it is best described in terms of a plot of 
$m$ as a function of $r_+$. This is shown in figure 4, where we have 
given a plot of $m(r_+)$ for different $l$, keeping the other parameters fixed. 
For small $l$, $m$ monotonically decreases with $r_+$. So, the 
configuration has only one horizon (inner or cosmological depending on 
the mass).
However, for large $l$, the solution has three horizons; out of them the 
largest one is the cosmological 
horizon and the other two are the black hole inner horizon and the event 
horizon. For fixed $l$, if 
we decrease $m$, these two horizons of the black hole come closer and they 
meet at a certain value of $m$. For even lower values of $m$, only the 
cosmological horizon exists. Now for even larger value of $l$ (shown in 
dashed line in the figure), there can be only two horizons. Out of
that, the larger one is the cosmological horizon. As we now increase $m$, 
the event and the cosmological horizons meet. Beyond that value of $m$ we 
get naked singularity (all horizons vanish) at the origin.
\begin{figure}[ht]
\epsfxsize=9cm
\centerline{\epsfbox{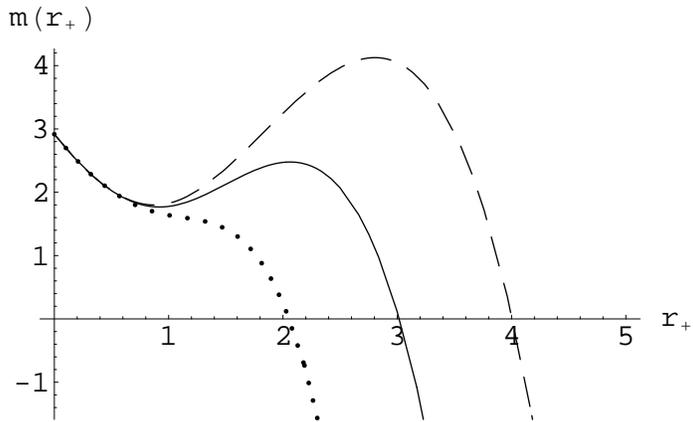}}
\caption{{\small{Mass $m$ as a function of $r_+$ for Born-Infeld deSitter holes 
for $n=4$, ${\beta} $=1, $q=1, l=2$ (dotted line), $3$ (solid line) , $4$  
(dashed line).}}}                             
\end{figure}

\section{Thermodynamics}
We now would like to study the thermodynamical properties of the
black holes we have just found.
   
\noindent The Hawking temperature of the hole can be calculated using the relation
\begin{equation}
{T}= {\frac{\kappa}{2\pi}} ,
\label{}
\end{equation}
where $ \kappa $ is the surface gravity and  is given by 
\begin{equation}
\kappa =
  {-\frac{1}{2}}{{d{g_{tt}}}\over{dr}}{\mid}_{r={r_+}} .
\end{equation}
$\kappa$ can be calculated explicitly from the metric functions. The 
temperature is then,
\begin{equation}
 T= {{\frac{1}{4\pi}}\Big[{\frac{n-2}{r_+}} +
 \Big\lbrace{\frac{4{\beta}^2}{n-1}}+\frac{n}{l^2} \Big 
  \rbrace{ r_+}-
{\frac{2{\sqrt2}\beta}{(n-1){r_+}^{n-2}}}{\sqrt{2{\beta}^2 
        {r_+}^{2n-2} + (n-1)(n-2)q^2}} \Big]}.
\label{temper}
\end{equation}
We note that when $r_+$ is such that the right hand side of the above 
equation is zero, the temperature of the black hole is zero. We notice 
that this gives the same constraint on $r_+$ that we have already 
encountered while discussing extremal black hole (see eq (\ref{extr})).
From here we conclude, in the extremal limit, the temperature of the
black hole is zero.\\ \\
\noindent Using the standard formula for entropy
\begin{equation}
 S = \int
{T^{-1}\Big({\frac{\partial{M}}{\partial{r_+}}}\Big)_{Q}dr_+},
\end{equation}
we get
\begin{equation}
{S} = 4 \pi \omega_{n-1} {{r_+}^{n-1}}
\label{}
\end{equation}
as the entropy of the black hole. It is indeed proportional to the area of 
the horizon. Now that we have all the relevant thermodynamic quantities, 
we can easily verify that the first law of thermodynamics. We find that 
\begin{equation}
dM = TdS +{\Phi}dQ, 
\end{equation}
is satisfied. 
To find the stability of the black hole, it is important to find the 
specific heat of the hole. This can be easily evaluated using 
\begin{equation}
C_Q = \Big({\partial M\over{\partial T}}\Big)_Q = 
\Big({\partial M\over{\partial r_+}}\Big)_Q/\Big({\partial T\over{\partial 
r_+}}\Big)_Q,
\end{equation}
where $M$ and $T$ are given in eqs. (\ref{mas}) and (\ref{temper}). In 
figure 5, we have plotted $C_Q$ as a 
\begin{figure}[ht]
\epsfxsize=8cm
\centerline{\epsfbox{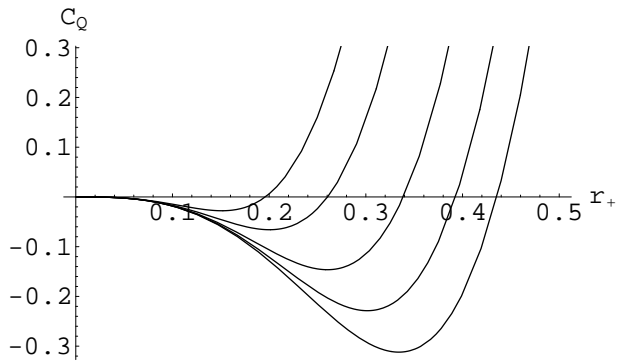}}
\caption{{\small{Specific heat at fixed charge as a 
function of $r_+$ where $n=4, l=.2, \beta=1$ and  $q=.5, 1, 2, 3, 4$
from left to right respectively.  }}}  
\end{figure}
\\
function of horizon radius $r_+$.  We see from the figure that the
specific heat is negative if $r_+$ is less than certain value; making the 
hole 
{\it unstable}. We therefore conclude that the large Born-Infeld black 
holes with adS asymptotics are stable against fluctuations.
We can now write down the thermodynamics canonical potential  at {\it 
fixed} charge $Q$, 
$ F = E-TS $,
where $E = M - M_e$ and $ M_e$ is the ADM mass of the extremal black hole
that follows from the condition (\ref{extr}). We get
\begin{eqnarray}
F &=&{\omega}_{n-1}\Big[ {r_+}^{n-2}
-\Big\lbrace\frac{4{\beta}^2}{n(n-1)}+\frac{1}{l^2} \Big \rbrace{r_+}^n 
+{{2\sqrt{2}\beta
r_+}\over{n(n-1)}}{\sqrt{ 2 {\beta}^2 
{r_+}^{2n-2}+(n-1)(n-2)q^2}}\nonumber \\
 &+& {
{2(n-1)^2q^2}\over{ n r_+^{n-2}}}{{}_2F_1}{
\Big[ {{n-2}\over{2n-2}}, {1\over 2}, {3n-4\over{2n-2}}, 
-{{(n-1)(n-2)q^2}\over{2 \beta^2 {r_+}^{2n-2}}}\Big]}-M_e\Big].
\label{}
\end{eqnarray}
In figure 6, we have shown $F$ as a function of $r_+$ for Born-Infeld 
black holes and for the standard RNadS black holes. This figure also shows 
that the larger black holes are more stable than the smaller ones. We
also see that for large $ r_+ $, $ F $ behaves similar to that of RNadS 
black holes. However, for small holes, their behaviours are distinctly 
different.

\begin{figure}
\epsfxsize=8cm
\centerline{\epsfbox{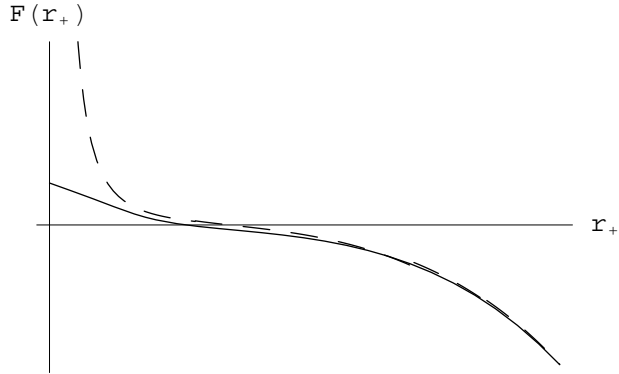}}
\caption{{\small{Free energy at fixed charge for Born-Infeld (solid line) and 
RNadS (dashed line) black holes for $n =4$, $l=1$,$ \beta=10$ and $q=10$.
}}}                             
\end{figure}

\section{Discussion }

In this letter, we have constructed charged adS and dS black holes of 
Einstein-Born-Infeld actions. We believe that this is only the first step 
before we analyse, following the adS/CFT correspondence, the behaviour of 
the boundary theory caused by higher order gauge field perturbations in 
the bulk. Beside the motivation coming from adS/CFT side, Born-Infeld 
lagrangian appears very frequently in string theory. So we expect that  
it is worthwhile to know various properties of black hole 
solutions in this theory.

\noindent Though, in this paper we have constructed the Born-Infeld black 
holes in the presence of a cosmological constant and discussed their 
thermodynamical properties,  many issues however still remain to be 
investigated. We know that 
Reissner-Nordstrom adS black holes undergo Hawking-Page phase 
transition. This transition gets modified as we include Born-Infeld 
corrections into account. We hope to carry out a detail study on this issue 
in the future. Furthermore, in the context of brane world cosmology, 
it was found that a brane moving in a Reisssner-Nordstrom adS background 
generates non-singular cosmology \cite{bpm}. However, as 
shown in 
\cite{hm}, the brane always crosses the inner horizon of the bulk 
geometry, creating instability. It would be interesting to study cosmology 
on the brane when it is moving in the charged black hole backgrounds that 
we have constructed. Note that since these charged holes does not have 
inner 
horizon for certain range of parameters, we may generate non-singular 
cosmology without creating the instabilities that we have just mentioned.

\bigskip

\noindent{\large\bf{Acknowledgements:}}

\noindent I would like to thank S. Mukherji for suggesting me this
problem, for fruitful  discussions at every step of the work and for
help in structuring the manuscript. I would also like to thank A. Biswas for
several useful conversations.


\begin{thebibliography}{99}

\bibitem{review} O. Aharony, S. Gubser, J. Maldacena, H.Ooguri and Y. Oz,
Phys. Rept. 323 (2000) 183, hep-th/9905111.

\bibitem{gkt} S. Gubser, I. Klebanov and A. Tseytlin, Nucl. Phys. {\bf 
B534} 
(1998) 202,
hep-th/9805156.

\bibitem{li} M. Li, JHEP 9903 (1999) 004, hep-th/9807196.

\bibitem{lindstrom} K. Landsteiner, Mod. Phys. Lett. {\bf A14} (1999) 379, 
hep-th/9901143.

\bibitem{mozo} M. Vazquez-Mozo, Phys. Rev. {\bf D60} (1999) 106010, 
hep-th/9905030.

\bibitem{myers} C. Burgess, N. Constable, R. Myers, JHEP 9908 (1999) 017, 
hep-th/9907188.

\bibitem{nod}M. Cvetic, S. Nojiri, S. Odintsov, Nucl. Phys. {\bf B628} 
(2002) 295, hep-th/0112045;
S. Nojiri and S. Odintsov, Phys. Rev. {\bf D66} (2002) 044012,
hep-th/0204112.

\bibitem{banados} M. Banados, Phys. Lett. {\bf B579} (2004) 13, 
hep-th/0310160.

\bibitem{emparan} A. Chamblin, R. Emparan, C. Johnson and R. Myers, Phys.
Rev. {\bf D60} (1999) 064018, hep-th/9902170.

\bibitem{cg} M. Cataldo and A. Garcia, Phys. Lett. {\bf B456} (1999) 28, 
hep-th/9903257.

\bibitem{fk} S. Fernando and D. Kung, Gen. Rel. Grav. {\bf 35} (2003) 129,
hep-th/0306120.

\bibitem{rasheed} D. Rasheed, "Non-linear electrodynamics: zeroth and 
first laws of black hole thermodynamics", hep-th/9702087.

\bibitem{emp} A. Chamblin, R. Emparan, C. Johnson and R. Myers, Phys. 
Rev. {\bf D60} (1999) 104026, hep-th/9904197.

\bibitem{bpm} A.K. Biswas and S. Mukherji, JHEP {\bf 0103} (2001) 046,
hep-th/0102138;
S. Mukherji and M. Peloso, Phys. Lett. {\bf B 547} (2002) 297, 
hep-th/0205180;
A. Biswas, S. Mukherji and S.S. Pal, Int. Jour. Mod. Phys. {\bf A19} 
(2004) 557, hep-th/0301144.

\bibitem{hm} J.L. Hovdebo and R.C. Myers, JCAP {\bf 0311} (2003) 012, 
hepth/0308088.

\end{thebibliography}
\end{document}